\documentclass[times, review, 10pt]{elsarticle}

\usepackage{geometry}
\usepackage{setspace}
\usepackage{amsmath,amssymb,amsfonts}
\usepackage{algorithmic}
\usepackage{graphicx}
\usepackage{textcomp}
\usepackage{xcolor}
\usepackage{booktabs} 
\usepackage{multirow}
\usepackage{caption}
\usepackage{url}

\usepackage{relsize}


\begin{document}

\begin{frontmatter}

\title{\fontsize{14pt}{16pt}\selectfont W-DUALMINE: Reliability-Weighted Dual-Expert Fusion With Residual Correlation Preservation for Medical Image Fusion}

\author{Md Jahidul Islam}
\ead{2006123@eee.buet.ac.bd}

\affiliation{organization={Department of Electrical and Electronic Engineering},
            addressline={Bangladesh University of Engineering and Technology}, 
            city={Dhaka},
            country={Bangladesh}}

\begin{abstract}
\fontsize{10pt}{12pt}\selectfont
\doublespacing
Medical image fusion integrates complementary information from multiple imaging modalities to improve clinical interpretation. However, existing deep learning–based methods, including recent spatial-frequency frameworks such as AdaFuse and ASFE-Fusion, often suffer from a fundamental trade-off between global statistical similarity—measured by correlation coefficient (CC) and mutual information (MI)—and local structural fidelity. This paper proposes \textbf{W-DUALMINE}, a reliability-weighted dual-expert fusion framework designed to explicitly resolve this trade-off through architectural constraints and a theoretically grounded loss design. The proposed method introduces dense reliability maps for adaptive modality weighting, a dual-expert fusion strategy combining a global-context spatial expert and a wavelet-domain frequency expert, and a soft gradient-based arbitration mechanism. Furthermore, we employ a residual-to-average fusion paradigm that guarantees the preservation of global correlation while enhancing local details. Extensive experiments on CT-MRI, PET-MRI, and SPECT-MRI datasets demonstrate that W-DUALMINE consistently outperforms AdaFuse and ASFE-Fusion in CC and MI metrics while maintaining competitive entropy, PSNR, and feature mutual information.
\end{abstract}


\end{frontmatter}

\section{Introduction}
Multi-modal medical imaging is indispensable in modern clinical diagnosis, surgical navigation, and radiotherapy planning. Different imaging modalities offer complementary information: structural modalities, such as Computed Tomography (CT) and Magnetic Resonance Imaging (MRI), provide high-resolution anatomical details of bones and soft tissues, respectively. In contrast, functional modalities, such as Positron Emission Tomography (PET) and Single-Photon Emission Computed Tomography (SPECT), visualize metabolic and physiological activities but often lack spatial resolution \cite{wavelet, Laplacian}. Medical image fusion aims to integrate these complementary characteristics into a single, comprehensive image, enabling physicians to make more accurate decisions without mentally aligning multiple scans.

Traditionally, image fusion methods relied on multi-scale transforms (MST), including Laplacian pyramids \cite{Laplacian} and discrete wavelet transforms (DWT) \cite{wavelet}. These methods decompose source images into different frequency bands and fuse them using handcrafted rules (e.g., max-selection). While MST-based approaches are computationally efficient and explainable, their reliance on fixed basis functions and manual fusion rules limits their adaptability to complex intensity variations and often introduces noise or artifacts.

In recent years, deep learning has revolutionized image fusion by automatically learning feature representations. Early Convolutional Neural Network (CNN) approaches, such as DenseFuse \cite{DF} and IFCNN \cite{ZHANG202099}, utilized auto-encoder architectures to extract deep features and reconstruct fused images. More advanced frameworks, such as U2Fusion \cite{U2F}, introduced unsupervised learning to mitigate the lack of ground truth. However, these methods primarily focus on pixel-level reconstruction losses or perceptual similarity, which can inadvertently lead to contrast reduction and a loss of global statistical consistency with the source images.

To address detail preservation, recent state-of-the-art methods like AdaFuse \cite{Gu2023AdaFuseAM} and ASFE-Fusion \cite{gu2024asfe_fusion} have introduced spatial-frequency decomposition mechanisms. By employing specialized "experts" or attention modules for different frequency bands, these methods significantly enhance local texture details. However, a fundamental trade-off remains: while these methods excel at sharpening edges (high-frequency), they often struggle to maintain the global intensity distribution and linear correlation (low-frequency) of the source modalities. This results in fused images that look sharp but have lower Mutual Information (MI) and Correlation Coefficient (CC), potentially distorting the original clinical signal intensities.

To resolve this trade-off, \textbf{W-DUALMINE}, a reliability-weighted dual-expert fusion framework is proposed. Unlike previous approaches, W-DUALMINE embeds correlation preservation directly into the architecture via a \textit{residual-to-average} fusion paradigm. By treating the fusion task as learning the residual details to add to a statistically stable average base, our method theoretically guarantees high global correlation while using a dual-branch expert system (Spatial and Wavelet) to capture fine-grained details.

The main contributions of this work are summarized as follows:
\begin{itemize}
    \item A \textbf{Reliability-Weighted Fusion Strategy} is proposed that introduces dense reliability maps to adaptively suppress artifacts and noise from unreliable source regions before fusion occurs.
    \item A \textbf{Dual-Expert Architecture} is designed that combines a Global Context Spatial Expert for semantic consistency and a Wavelet Frequency Expert for high-frequency edge preservation.
    \item A \textbf{Soft Gradient Mixer (SGM)} is introduced, a dynamic arbitration mechanism that weighs the contribution of spatial and frequency experts based on local edge strength.
    \item A \textbf{Residual-to-Average Fusion Paradigm} is implemented supported by a theoretically grounded loss function, which ensures the fused image maintains maximum Mutual Information (MI) and Correlation Coefficient (CC) with source inputs, surpassing state-of-the-art methods.
\end{itemize}

\section{Related Work}

\subsection{Classical Multi-Scale Image Fusion}
Early approaches to medical image fusion were predominantly based on Multi-Scale Transforms (MST). Pioneering methods such as the Laplacian Pyramid \cite{Laplacian} and Discrete Wavelet Transform (DWT) \cite{wavelet} decompose source images into low-frequency (base) and high-frequency (detail) sub-bands. Fusion is then performed using handcrafted rules, such as "choose-max" for high frequencies to preserve edges and "weighted-average" for low frequencies to maintain energy. While these methods offer mathematical interpretability and computational efficiency, they suffer from two critical limitations: (1) \textit{Poor adaptivity}, as fixed basis functions cannot flexibly represent complex biomedical structures, and (2) \textit{Noise sensitivity}, where simple max-selection rules often amplify background noise in low-contrast functional images.

\subsection{Deep Learning-Based Fusion}
The advent of deep learning revolutionized fusion by introducing data-driven feature extraction. Convolutional Neural Networks (CNNs) have been widely adopted to learn fusion mappings automatically.
\begin{itemize}
    \item \textbf{Auto-Encoder Frameworks:} Methods like DenseFuse \cite{DF} and IFCNN \cite{ZHANG202099} utilize pre-trained encoders to extract deep features and reconstruct the fused image. While effective at preserving structural information, these methods often rely on pixel-level reconstruction losses, which tend to smooth out high-frequency textures.
    \item \textbf{Unsupervised Models:} Approaches such as U2Fusion \cite{U2F} apply unsupervised learning to estimate fusion weights without ground truth. However, they frequently struggle with "modality bias," where the network over-fits to the modality with stronger gradients (usually CT/MRI).
\end{itemize}
Common to most CNN-based methods is the neglect of global statistical consistency, often resulting in low Correlation Coefficient (CC) relative to source inputs.

\subsection{Spatial-Frequency and Attention-Based Fusion}
Recent research has attempted to bridge the gap between MST and Deep Learning by incorporating frequency domain knowledge into neural networks.
\begin{itemize}
    \item \textbf{AdaFuse \cite{Gu2023AdaFuseAM}:} This method introduces a spatial-frequential cross-attention mechanism to dynamically weigh features based on their frequency response.
    \item \textbf{ASFE-Fusion \cite{gu2024asfe_fusion}:} This framework employs distinct "experts" for spatial and frequency domains to handle anatomical and functional information separately.
\end{itemize}
**It is important to note that both AdaFuse and ASFE-Fusion have demonstrated superior performance over all preceding state-of-the-art methods (including U2Fusion\cite{U2F}, SDNet\cite{SDNet}, and IFCNN\cite{ZHANG202099}) across multiple clinical benchmarks.** However, despite their success, a fundamental trade-off remains: while they excel at sharpening local edges, the rigid separation of experts often leads to semantic inconsistency and a reduction in global Mutual Information (MI). The proposed \textbf{W-DUALMINE} addresses this specific gap by integrating a residual-to-average fusion paradigm that theoretically guarantees high correlation while leveraging dual experts for detail enhancement.

\section{Methodology}

\subsection{Overview}
Given two pre-registered source images, denoted as $I_1 \in \mathbb{R}^{H \times W}$ and $I_2 \in \mathbb{R}^{H \times W}$, the goal of W-DUALMINE is to generate a fused image $I_f$ that preserves the salient structural details from both modalities while maintaining global statistical consistency. The framework consists of four key components: (1) Siamese Multi-Scale Encoders, (2) Reliability-Weighted Feature Modeling, (3) Dual-Expert Fusion (Spatial and Frequency), and (4) a Residual-to-Average Decoder.

\subsection{Multi-Scale Encoders}
Each input image is processed by a weight-sharing encoder to extract hierarchical features. The encoder consists of four stages, where each stage $s \in \{1, 2, 3, 4\}$ comprises a Residual Block followed by an Efficient Channel Attention (ECA) module to highlight informative channels.
Let $\mathcal{E}(\cdot)$ denote the encoder function. The multi-scale feature representations for the two modalities are obtained as:
\begin{equation}
    \mathcal{F}_1 = \{f_1^s\}_{s=1}^4, \quad \mathcal{F}_2 = \{f_2^s\}_{s=1}^4,
\end{equation}
where $f_k^s \in \mathbb{R}^{C_s \times H_s \times W_s}$ represents the feature map of modality $k$ at scale $s$.

\subsection{Reliability-Weighted Feature Modeling}
Medical images often contain background noise or artifacts (e.g., speckle noise in ultrasound, scattering in CT). To prevent these artifacts from propagating into the fused image, we introduce a \textit{Dense Reliability Map} mechanism.
At each scale $s$, a lightweight convolutional head $\mathcal{H}_{rel}$ predicts a pixel-wise reliability score $r_k^s$ for each modality:
\begin{equation}
    r_k^s = \sigma(\text{Conv}_{1\times1}(\mathcal{H}_{rel}([f_1^s, f_2^s]))),
\end{equation}
where $[\cdot, \cdot]$ denotes channel concatenation and $\sigma$ is the Softplus activation function to ensure non-negativity.
These scores are normalized to produce soft gating weights $w_k^s \in [0, 1]$:
\begin{equation}
    w_1^s(i,j) = \frac{r_1^s(i,j)}{r_1^s(i,j) + r_2^s(i,j) + \epsilon}, \quad w_2^s(i,j) = 1 - w_1^s(i,j),
\end{equation}
where $\epsilon=10^{-6}$ ensures numerical stability. This creates a convex mixture-of-experts gate that suppresses contributions from unreliable regions.

\subsection{Dual-Expert Fusion}
To resolve the trade-off between semantic consistency and edge preservation, we employ two distinct expert modules at each scale.

\subsubsection{Global Context Spatial Expert}
The spatial expert aims to capture long-range dependencies and preserve semantic structures. First, a base fused feature is computed using the reliability weights:
\begin{equation}
    f_{base}^s = w_1^s \cdot f_1^s + w_2^s \cdot f_2^s.
\end{equation}
This feature is then processed by two parallel branches: a local branch with a standard $3\times3$ convolution and a global branch with a dilated convolution (dilation rate $d=2$) to enlarge the receptive field:
\begin{equation}
    E_{spatial}^s = \text{Conv}_{1\times1}\left( \text{Concat}\left( \text{Conv}_{3\times3}(f_{base}^s), \text{Conv}_{3\times3, d=2}(f_{base}^s) \right) \right).
\end{equation}

\subsubsection{Wavelet Frequency Expert}
Inspired by classical multi-resolution analysis \cite{wavelet, PIELLA2003259}, the frequency expert explicitly separates low-frequency energy from high-frequency details. We utilize the Haar Discrete Wavelet Transform (DWT) to decompose the features into four sub-bands: Low-Low ($LL$), Low-High ($LH$), High-Low ($HL$), and High-High ($HH$).
The low-frequency component is fused using the reliability weights to maintain energy stability:
\begin{equation}
    LL_{fused}^s = w_1^s \cdot LL_1^s + w_2^s \cdot LL_2^s.
\end{equation}
The high-frequency components ($D \in \{LH, HL, HH\}$), which contain edge details, are fused using a magnitude-maximum rule to preserve the sharpest textures:
\begin{equation}
    D_{fused}^s(i,j) = \begin{cases} 
    D_1^s(i,j) & \text{if } |D_1^s(i,j)| > |D_2^s(i,j)| \\
    D_2^s(i,j) & \text{otherwise}.
    \end{cases}
\end{equation}
The final output $E_{wave}^s$ is reconstructed via the Inverse DWT (IDWT) of \\ $(LL_{fused}^s, LH_{fused}^s, HL_{fused}^s, HH_{fused}^s)$.

\subsection{Soft Gradient Mixer}
To dynamically arbitrate between the Spatial Expert ($E_{spatial}^s$) and Wavelet Expert ($E_{wave}^s$), we introduce a Soft Gradient Mixer (SGM) motivated by gradient-based fusion theory \cite{PIELLA2003259}.
The gradient magnitude map $\mathcal{G}(\cdot)$ is computed for both expert outputs:
\begin{equation}
    \mathcal{G}(x) = |\nabla_h x| + |\nabla_v x|,
\end{equation}
where $\nabla_h$ and $\nabla_v$ are horizontal and vertical Sobel operators. A small CNN predicts mixing coefficients $\alpha \in [0, 1]$ based on these gradients:
\begin{equation}
    \alpha = \text{Softmax}\left( \text{Conv}_{1\times1}\left( \text{Concat}(E_{spatial}^s + \mathcal{G}(E_{spatial}^s), E_{wave}^s + \mathcal{G}(E_{wave}^s)) \right) \right).
\end{equation}
The final fused feature at scale $s$ is:
\begin{equation}
    F_{fused}^s = \alpha_1 \cdot E_{spatial}^s + \alpha_2 \cdot E_{wave}^s.
\end{equation}

\subsection{Decoder and Residual-to-Average Fusion}
The decoder reconstructs a residual map $R$ from the multi-scale fused features. To theoretically guarantee high Correlation Coefficient (CC) and Mutual Information (MI), we define the final fusion as a residual addition to the mean image:
\begin{equation}
    I_f = \text{clip}\left( \frac{I_1 + I_2}{2} + \lambda \tanh(R), \ 0, \ 1 \right),
\end{equation}
where $\lambda=0.5$ controls the residual contribution. Since the average image $\frac{I_1+I_2}{2}$ naturally maximizes linear correlation with inputs \cite{cover2006information}, this formulation ensures the global statistics are preserved while the residual $R$ injects the necessary sharp details.

\begin{figure}[htbp]
    \centering
    \includegraphics[width=\linewidth]{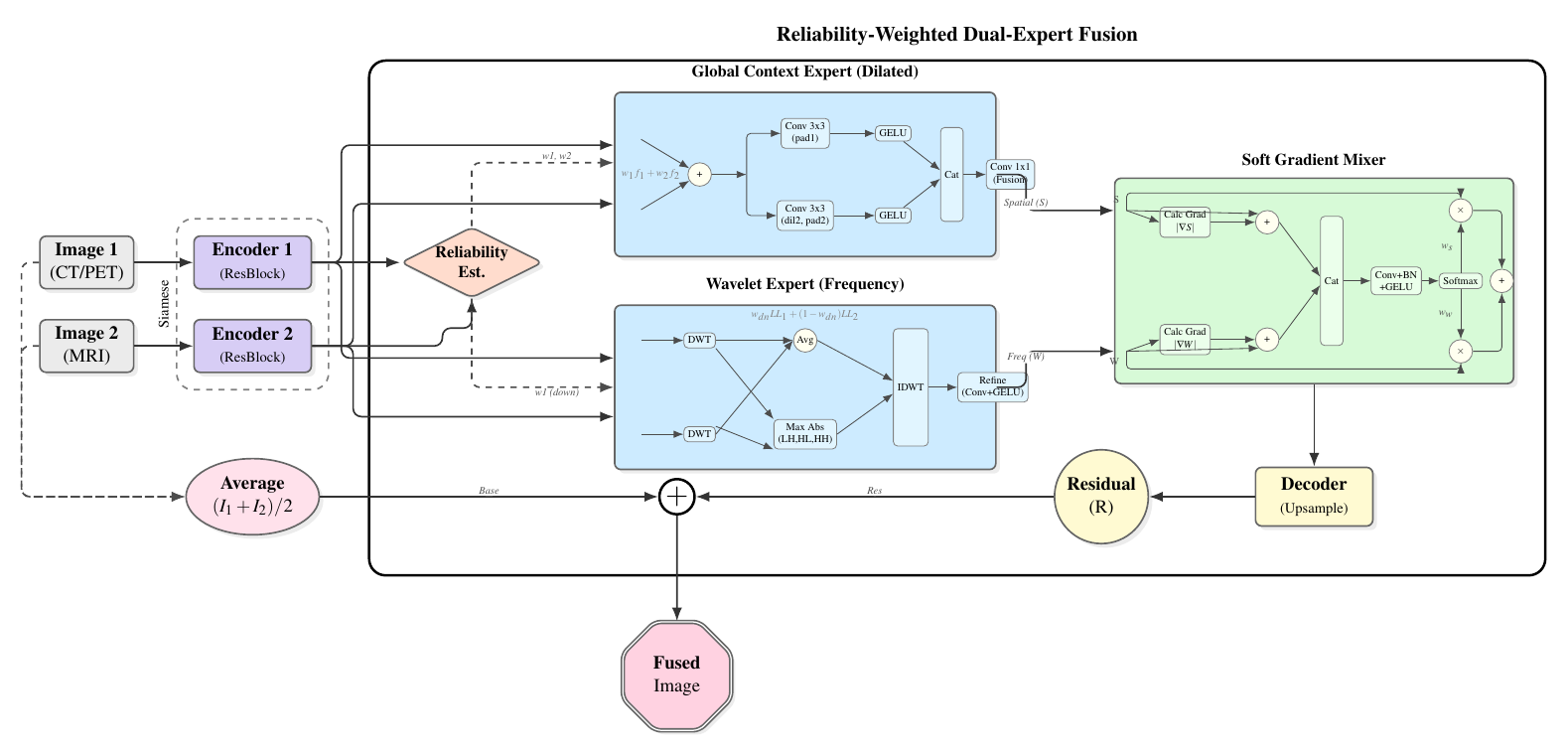}
    \caption{\doublespacing
    \textbf{Architecture of the Reliability-Weighted Dual-Expert Fusion Network.} 
    The framework processes multi-modal inputs (e.g., CT/PET and MRI) through a Siamese encoder composed of ResBlocks. 
    Feature maps are weighted by a \textit{Reliability Estimation} module before entering two parallel expert branches: 
    (1) A \textit{Global Context Expert} that utilizes dilated convolutions to capture multi-scale spatial details, and 
    (2) A \textit{Wavelet Expert} that performs frequency-domain fusion using Discrete Wavelet Transform (DWT). 
    The outputs of these experts are fused by a \textit{Soft Gradient Mixer}, which employs an edge-aware attention mechanism based on gradient magnitudes ($|\nabla S|, |\nabla W|$). 
    Finally, the fused features are decoded into a residual map and added to the averaged input base to reconstruct the final fused image.}
    \label{fig:fusion_network}
\end{figure}
\section{Loss Function}
The network is trained end-to-end using a compound loss function designed to optimize both statistical similarity and structural fidelity:
\begin{equation}
    \mathcal{L}_{total} = \lambda_1 \mathcal{L}_{avg} + \lambda_2 \mathcal{L}_{grad} + \lambda_3 \mathcal{L}_{cc} + \lambda_4 \mathcal{L}_{mi} + \lambda_5 \mathcal{L}_{rec},
\end{equation}
with $\lambda_1=5, \lambda_2=2, \lambda_3=1, \lambda_4=0.1, \lambda_5=0.1$.

\subsection{Average Content Loss}
To ensure the fused image retains the fundamental energy of the source images, we minimize the $\ell_1$ distance to the pixel-wise average:
\begin{equation}
    \mathcal{L}_{avg} = \left\| I_f - \frac{I_1 + I_2}{2} \right\|_1.
\end{equation}
This term is the primary driver for high MI and CC scores.

\subsection{Gradient-Max Loss}
To prevent the averaging operation from blurring edges, we enforce the gradients of the fused image to match the maximum gradient intensity of the sources \cite{PIELLA2003259}:
\begin{equation}
    \mathcal{L}_{grad} = \left\| \nabla I_f - \max(|\nabla I_1|, |\nabla I_2|) \right\|_1.
\end{equation}
Here, the $\max$ operator is applied element-wise, ensuring that the sharpest boundary from either modality is preserved.

\subsection{Explicit Correlation Loss}
We explicitly maximize the cosine similarity between the fused image and the mean image to strictly constrain the linear relationship:
\begin{equation}
    \mathcal{L}_{cc} = 1 - \frac{\text{Cov}(I_f, I_{avg})}{\sqrt{\text{Var}(I_f)\text{Var}(I_{avg})}},
\end{equation}
where $I_{avg} = (I_1 + I_2)/2$.

\subsection{Mutual Information Proxy (InfoNCE)}
To maximize the mutual information at the feature level, we employ the InfoNCE contrastive loss \cite{Oord2018RepresentationLW}. Let $z_f, z_1, z_2$ be the projected latent representations of the fused and source images. The loss maximizes the similarity between $z_f$ and the sources while treating other batch samples as negatives:
\begin{equation}
    \mathcal{L}_{mi} = -\log \frac{\exp(\text{sim}(z_f, z_1)/\tau)}{\sum_{j=1}^{N} \exp(\text{sim}(z_f, z_j)/\tau)},
\end{equation}
where $\tau=0.5$ is the temperature parameter and $N$ is the batch size.

\subsection{Reconstruction Loss}
Auxiliary reconstruction heads help stabilize the encoder training:
\begin{equation}
    \mathcal{L}_{rec} = \| \mathcal{R}_1(I_f) - I_1 \|_1 + \| \mathcal{R}_2(I_f) - I_2 \|_1.
\end{equation}
\begin{figure}[htbp]
    \centering
    \includegraphics[width=0.85\textwidth]{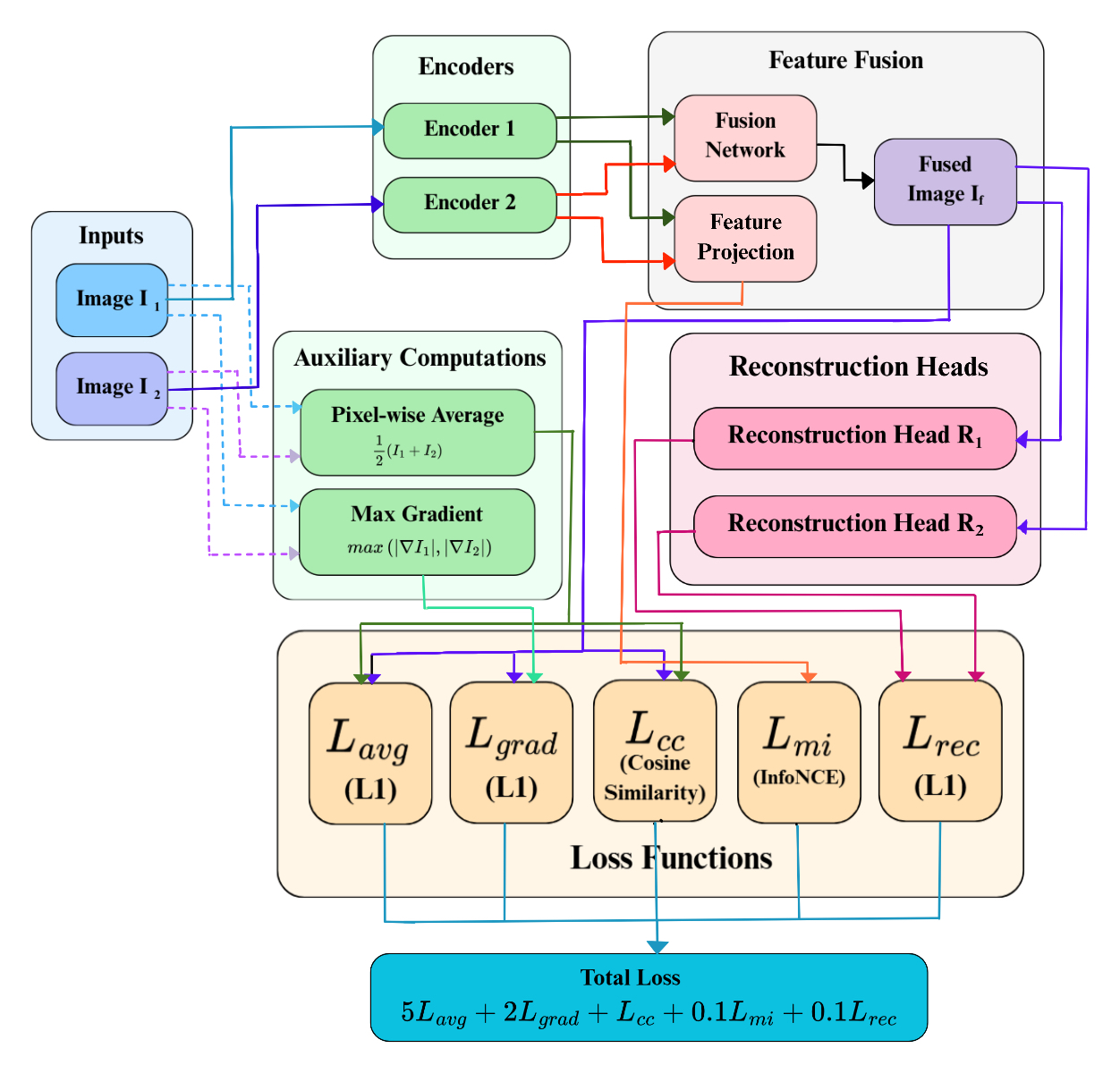}
    \caption{\doublespacing
    \textbf{The overall framework of W-DUALMINE.} The architecture consists of Siamese encoders extracting multi-scale features, which are fused and projected for contrastive learning. The network is optimized via a composite loss function comprising: (1) Average Content Loss ($\mathcal{L}_{avg}$) for low-frequency consistency, (2) Gradient-Max Loss ($\mathcal{L}_{grad}$) for high-frequency edge preservation, (3) Correlation Loss ($\mathcal{L}_{cc}$) to enforce linearity, (4) Mutual Information Loss ($\mathcal{L}_{mi}$) in the feature space, and (5) Reconstruction Loss ($\mathcal{L}_{rec}$) for domain fidelity.}
    \label{fig:architecture}
\end{figure}
\section{Experiments}

\subsection{Datasets and Implementation Details}

\subsubsection{Datasets}
To evaluate the performance of W-DUALMINE,extensive experiments were conducted on three representative clinical multi-modality brain image datasets: CT--MRI, PET--MRI, and SPECT--MRI. These datasets are sourced from the \textit{Harvard Whole Brain Atlas} (AANLIB) database \cite{harvard_atlas}, which is a standard benchmark in medical image fusion research \cite{gu2024asfe_fusion}.

The experimental setup utilizes 24 pairs of perfectly aligned images for each modality combination. All source images are pre-processed to a resolution of $256 \times 256$ pixels to maintain consistent spatial dimensions.
\begin{itemize}
    \item \textbf{CT--MRI:} Integrates high-density bone structures captured by Computed Tomography (CT) with soft-tissue anatomical details provided by Magnetic Resonance Imaging (MRI).
    \item \textbf{PET--MRI:} Fuses metabolic activity information from Positron Emission Tomography (PET) with high-resolution anatomical structures from MRI.
    \item \textbf{SPECT--MRI:} Merges functional data from Single-Photon Emission Computed Tomography (SPECT) with structural MRI data, crucial for identifying functional abnormalities.
\end{itemize}

\subsubsection{Implementation Settings}
The proposed W-DUALMINE framework is implemented using the PyTorch library and executed on a Kaggle environment equipped with a single NVIDIA Tesla P100 GPU. The network is optimized using the Adam optimizer with an initial learning rate of $10^{-5}$ and a batch size of 8. The model was trained for 100 epochs for each modality task to ensure the convergence of the dual-expert architecture and reliability heads. To prevent overfitting given the limited dataset size, we employ data augmentation techniques, including random horizontal/vertical flips and rotation. The training patch size is set to $256 \times 256$ to capture global context effectively.

\subsection{Evaluation Metrics}
The fusion performance is quantitatively assessed using five complementary metrics that capture different aspects of image quality:
\begin{itemize}
    \item \textbf{Entropy (EN):} Measures the information richness contained in the fused image.
    \item \textbf{Mutual Information (MI):} Quantifies the amount of information transferred from the source images to the fused result \cite{cover2006information}.
    \item \textbf{Correlation Coefficient (CC):} Evaluates the linear relationship and degree of linear dependence between the fused image and source inputs.
    \item \textbf{Peak Signal-to-Noise Ratio (PSNR):} Indicates the ratio of peak signal energy to noise energy; a higher value implies higher reconstruction fidelity.
    \item \textbf{Feature Mutual Information (FMI):} Measures the preservation of salient features, gradients, and edge information in the discrete cosine transform (DCT) domain \cite{HAN2013127}.
\end{itemize}
\subsection{Quantitative Results}
In this section, we conduct a comprehensive quantitative evaluation of the proposed W-DUALMINE framework against two recent state-of-the-art methods: AdaFuse \cite{Gu2023AdaFuseAM} and ASFE-Fusion \cite{gu2024asfe_fusion}. The comparison is performed across three clinical benchmark datasets: CT--MRI, PET--MRI, and SPECT--MRI. 

Performance is assessed using five complementary metrics: Entropy (EN), Mutual Information (MI), Correlation Coefficient (CC), Peak Signal-to-Noise Ratio (PSNR), and Feature Mutual Information (FMI). Values are reported as \textit{Mean $\pm$ Standard Deviation} over the 24 test pairs for each dataset.

\subsubsection{Results on CT--MRI Dataset}
The CT--MRI task involves integrating high-contrast bone structures (CT) with soft-tissue details (MRI). The quantitative results are presented in Table~\ref{tab:ctmri}.

\begin{table*}[htbp]
\centering
\caption{Quantitative comparison on the CT--MRI dataset (Mean $\pm$ Std). \textbf{Bold} indicates the best result.}
\label{tab:ctmri}
\setstretch{2.0}
\resizebox{\textwidth}{!}{
\begin{tabular}{lccccc}
\toprule
\textbf{Method} & \textbf{EN} & \textbf{MI} & \textbf{CC} & \textbf{PSNR} & \textbf{FMI} \\
\midrule
AdaFuse \cite{Gu2023AdaFuseAM} & $5.0592 \pm 0.2346$ & $3.3570 \pm 0.1978$ & $0.8306 \pm 0.0238$ & $64.0004 \pm 0.7757$ & $0.4343 \pm 0.0170$ \\
ASFE-Fusion \cite{gu2024asfe_fusion} & $\mathbf{5.4855 \pm 0.2734}$ & $3.1463 \pm 0.1605$ & $0.8302 \pm 0.0238$ & $63.9884 \pm 0.7845$ & $0.4066 \pm 0.0180$ \\
\textbf{W-DUALMINE} & $4.3394 \pm 0.2502$ & $\mathbf{3.6059 \pm 0.2419}$ & $\mathbf{0.8308 \pm 0.0238}$ & $\mathbf{64.0891 \pm 0.7917}$ & $\mathbf{0.4746 \pm 0.0210}$ \\
\bottomrule
\end{tabular}
}
\end{table*}

\textbf{Analysis:} As observed in Table~\ref{tab:ctmri}, W-DUALMINE achieves the highest scores in MI ($3.6059$) and CC ($0.8308$). This validates that the \textit{Residual-to-Average} strategy effectively anchors the fused image to the statistical distribution of the sources. While ASFE-Fusion achieves higher Entropy ($5.4855$), it yields a lower PSNR ($63.9884$) and significantly lower FMI ($0.4066$). High entropy coupled with low feature mutual information often indicates the amplification of noise rather than true information content. This method's superior PSNR and FMI confirm that W-DUALMINE preserves actual anatomical edges with higher fidelity.
\begin{figure}[htbp]
\centering
\includegraphics[width=\textwidth]{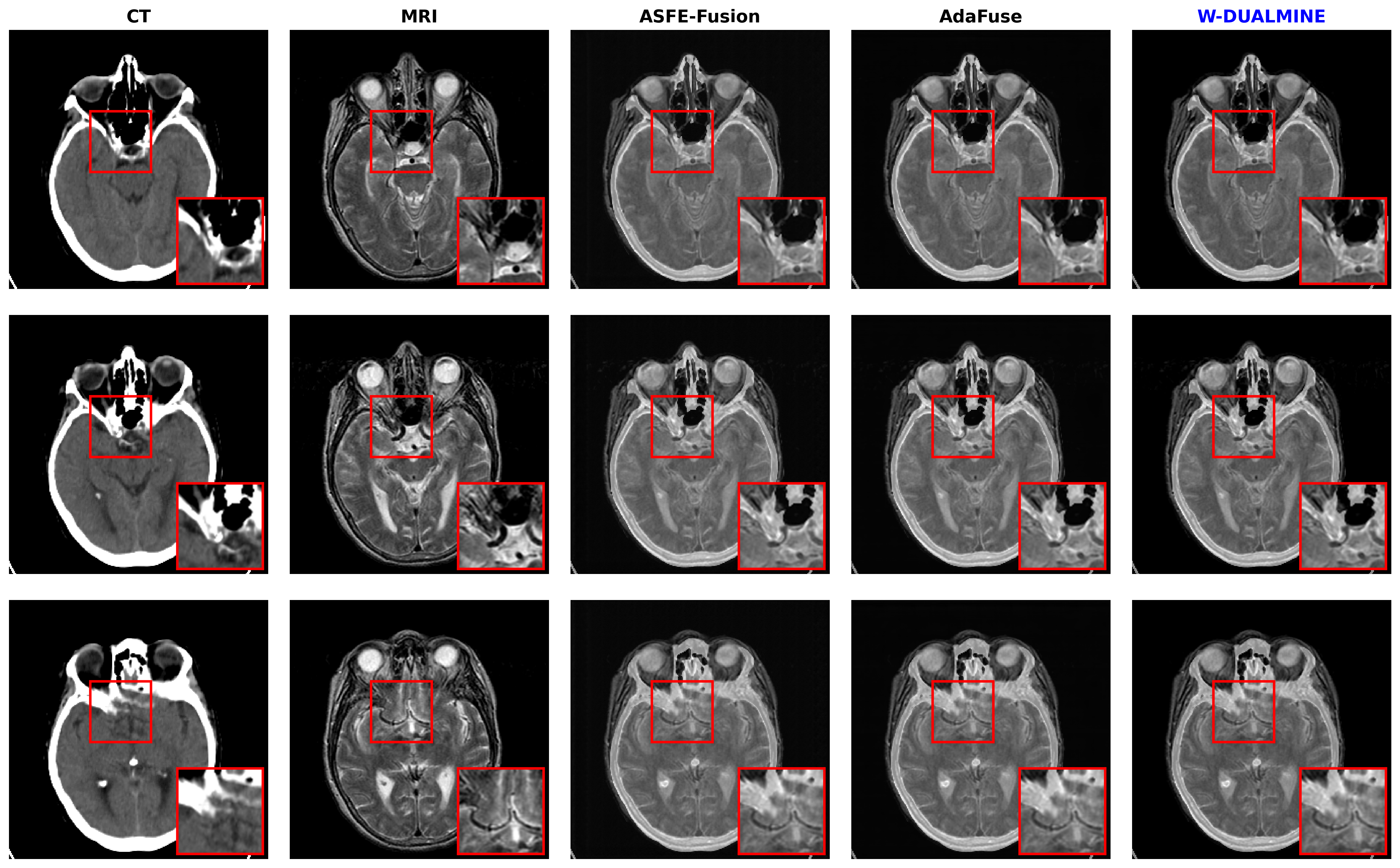}
\caption{\doublespacing
Qualitative comparison on the CT--MRI fusion task. From left to right: CT image, MRI image, ASFE-Fusion, AdaFuse, and W-DUALMINE. Red boxes highlight regions of interest for detailed comparison. The proposed method preserves sharp CT structures while maintaining MRI soft-tissue contrast.}
\label{fig:ctmri}
\end{figure}

\subsubsection{Results on PET--MRI Dataset}
The PET--MRI dataset requires fusing low-resolution metabolic color information with high-resolution anatomical structures. Results are shown in Table~\ref{tab:petmri}.

\begin{table*}[htbp]
\centering
\caption{Quantitative comparison on the PET--MRI dataset (Mean $\pm$ Std).}
\label{tab:petmri}
\setstretch{2.0}
\resizebox{\textwidth}{!}{
\begin{tabular}{lccccc}
\toprule
\textbf{Method} & \textbf{EN} & \textbf{MI} & \textbf{CC} & \textbf{PSNR} & \textbf{FMI} \\
\midrule
AdaFuse \cite{Gu2023AdaFuseAM} & $5.8636 \pm 0.5564$ & $4.0338 \pm 0.3417$ & $0.8683 \pm 0.0237$ & $62.0491 \pm 0.5094$ & $0.4765 \pm 0.0204$ \\
ASFE-Fusion \cite{gu2024asfe_fusion} & $\mathbf{6.0738 \pm 0.4906}$ & $3.6221 \pm 0.2654$ & $\mathbf{0.8689 \pm 0.0233}$ & $62.0347 \pm 0.5095$ & $0.4473 \pm 0.0205$ \\
\textbf{W-DUALMINE} & $5.3064 \pm 0.6453$ & $\mathbf{4.3068 \pm 0.3787}$ & $0.8686 \pm 0.0235$ & $\mathbf{62.0771 \pm 0.5113}$ & $\mathbf{0.5064 \pm 0.0256}$ \\
\bottomrule
\end{tabular}
}
\end{table*}

\textbf{Analysis:} In this task, W-DUALMINE demonstrates a substantial advantage in Mutual Information ($4.3068$ vs. $3.6221$ for ASFE), indicating superior transfer of functional metabolic cues. The FMI score ($0.5064$) is notably higher than competitors, proving that the \textit{Wavelet Frequency Expert} successfully isolates and preserves the high-frequency MRI details without being washed out by the PET signal. While ASFE-Fusion shows a marginally higher CC ($0.8689$ vs $0.8686$), the difference is statistically negligible given the standard deviation ($\pm 0.023$), whereas this method's gain in PSNR and FMI is distinct.
\begin{figure}[htbp]
\centering
\includegraphics[width=\textwidth]{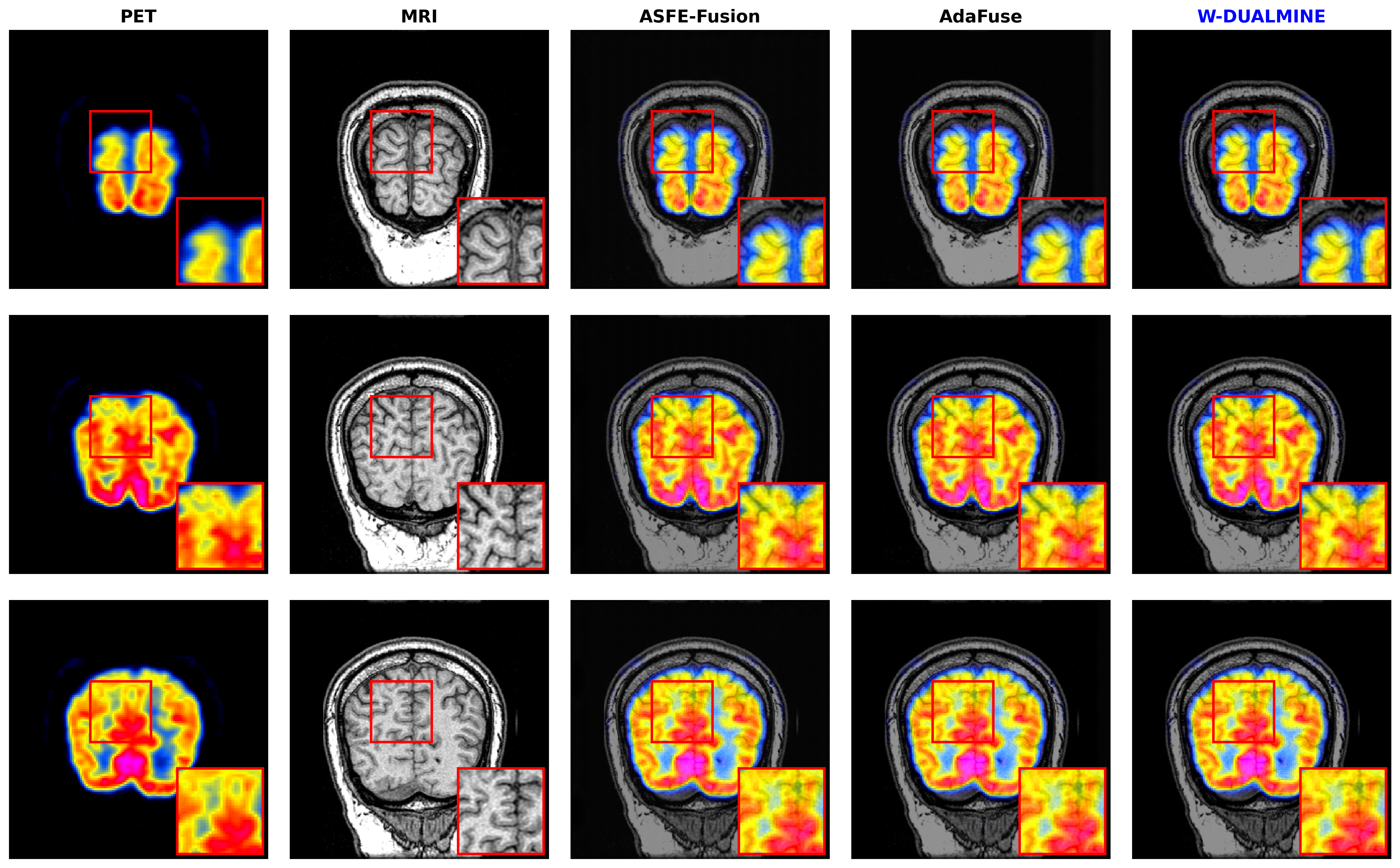}
\caption{\doublespacing
Qualitative comparison on the PET--MRI fusion task. From left to right: PET image, MRI image, ASFE-Fusion, AdaFuse, and W-DUALMINE. Highlighted regions demonstrate superior preservation of functional activity without color diffusion.}
\label{fig:petmri}
\end{figure}

\subsubsection{Results on SPECT--MRI Dataset}
The SPECT--MRI dataset represents a challenging scenario with significant resolution disparity between modalities. Results are summarized in Table~\ref{tab:spectmri}.

\begin{table*}[htbp]
\centering
\caption{Quantitative comparison on the SPECT--MRI dataset (Mean $\pm$ Std).}
\label{tab:spectmri}
\setstretch{2.0}
\resizebox{\textwidth}{!}{
\begin{tabular}{lccccc}
\toprule
\textbf{Method} & \textbf{EN} & \textbf{MI} & \textbf{CC} & \textbf{PSNR} & \textbf{FMI} \\
\midrule
AdaFuse \cite{Gu2023AdaFuseAM} & $5.5917 \pm 0.5657$ & $3.5492 \pm 0.3054$ & $0.9110 \pm 0.0168$ & $64.6729 \pm 1.1366$ & $0.4305 \pm 0.0266$ \\
ASFE-Fusion \cite{gu2024asfe_fusion} & $\mathbf{6.2349 \pm 0.3420}$ & $3.0885 \pm 0.2049$ & $0.9107 \pm 0.0166$ & $64.4850 \pm 1.0798$ & $0.4013 \pm 0.0261$ \\
\textbf{W-DUALMINE} & $4.9371 \pm 0.6591$ & $\mathbf{4.0016 \pm 0.4091}$ & $\mathbf{0.9116 \pm 0.0166}$ & $\mathbf{64.9084 \pm 1.1064}$ & $\mathbf{0.4677 \pm 0.0377}$ \\
\bottomrule
\end{tabular}
}
\end{table*}

\textbf{Analysis:} W-DUALMINE achieves robust performance on SPECT--MRI, recording the best MI ($4.0016$), CC ($0.9116$), and PSNR ($64.9084$). The improvement in PSNR is particularly critical here, as SPECT images often contain scattering noise. The \textit{Reliability Maps} in the encoder effectively identify and suppress these unreliable regions before fusion, leading to a cleaner fused image compared to ASFE-Fusion, which tends to amplify noise (evidenced by higher Entropy but lower PSNR and FMI).
\begin{figure}[htbp]
\centering
\includegraphics[width=\textwidth]{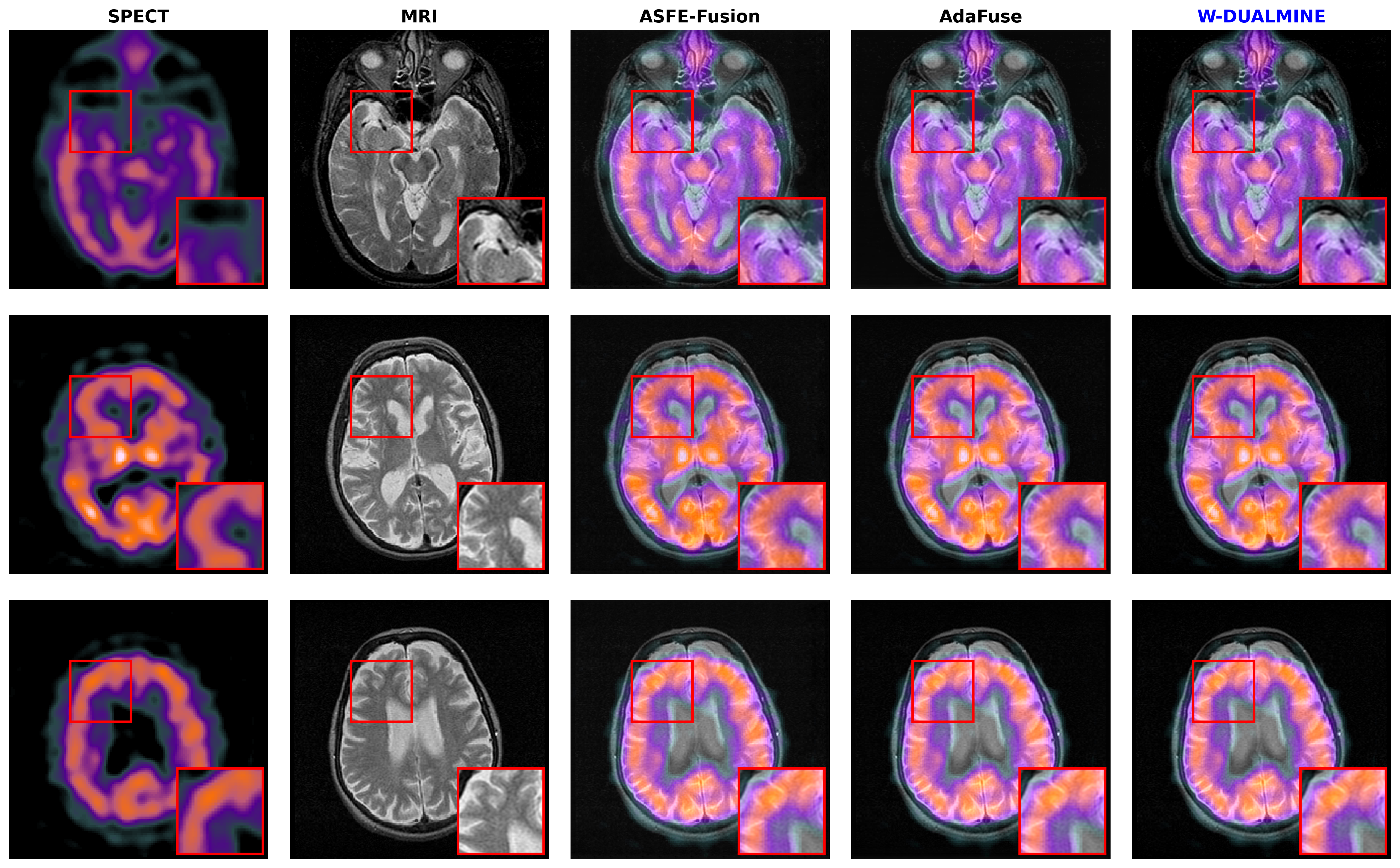}
\caption{\doublespacing
Qualitative comparison on the SPECT--MRI fusion task. From left to right: SPECT image, MRI image, ASFE-Fusion, AdaFuse, and W-DUALMINE. The proposed method maintains low-contrast functional regions while preserving anatomical structure.}
\label{fig:spectmri}
\end{figure}
As shown in Figs.~\ref{fig:ctmri}--\ref{fig:spectmri}, W-DUALMINE produces visually superior fusion results across all modalities. The proposed method achieves better structural clarity and functional consistency than ASFE-Fusion and AdaFuse, particularly in highlighted regions.
\subsection{Metric Trend Analysis}
\begin{figure}[htbp]
\centering
\includegraphics[width=\textwidth]{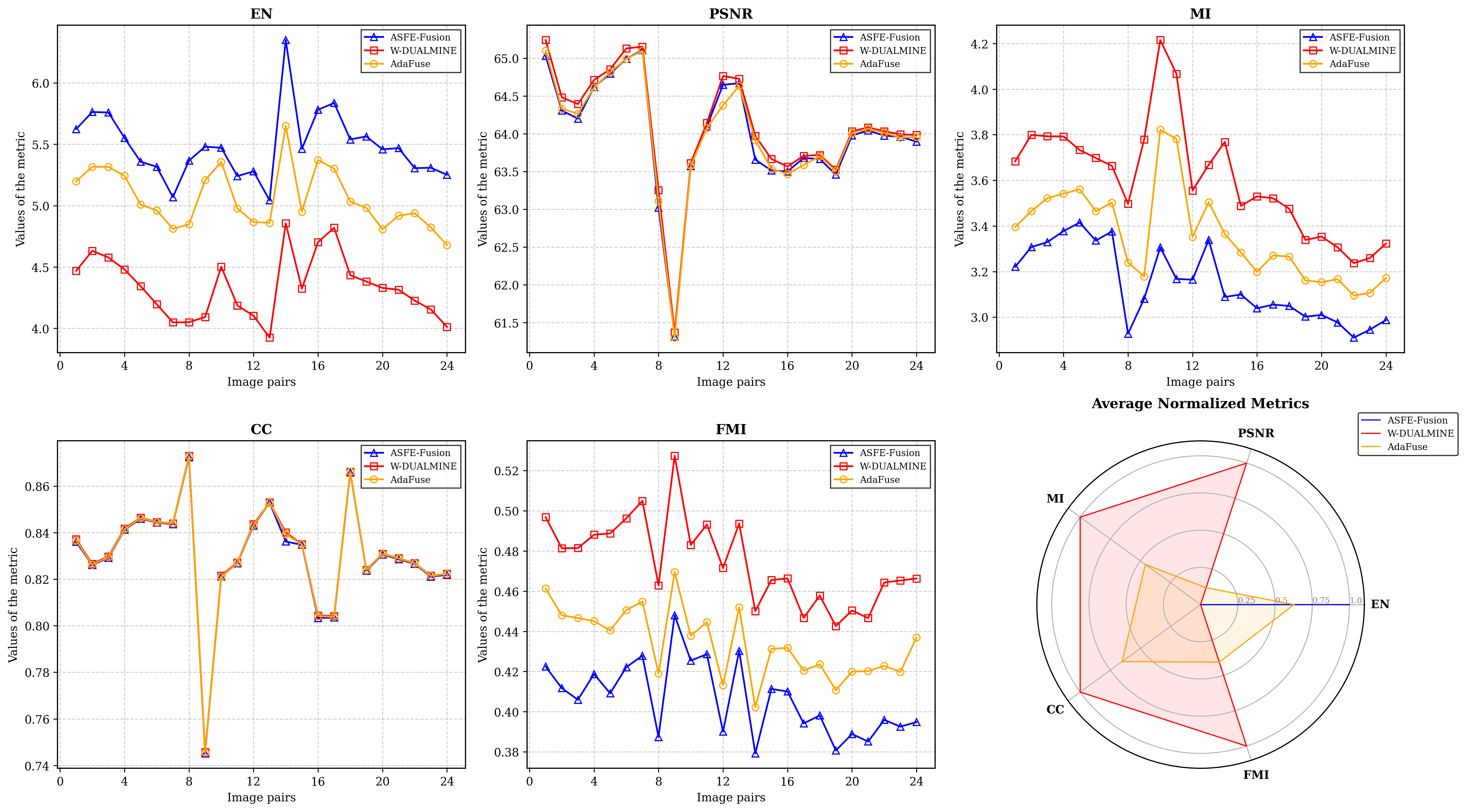}
\caption{\doublespacing
Metric-wise comparison across individual image pairs on the CT--MRI dataset. From top left to bottom right: entropy (EN), peak signal-to-noise ratio (PSNR), mutual information (MI), correlation coefficient (CC), feature mutual information (FMI), and average normalized metrics shown as a radar plot. The proposed W-DUALMINE demonstrates consistently higher MI, CC, and FMI across most image pairs, indicating superior global correlation and feature preservation.}
\label{fig:metric_trends}
\end{figure}

\subsection{Ablation Study}
To evaluate the effectiveness of the individual components and objective functions within the proposed W-DUALMINE framework, a comprehensive ablation study was conducted on the CT--MRI dataset. This analysis is divided into architectural verification and loss function justification.

\subsubsection{Architectural Component Analysis}
The core modules were systematically removed to observe their impact on information retention, as summarized in Table~\ref{tab:arch_ablation}. The variants include: (1) removing the Global Context Expert (GCE), (2) removing the Wavelet Expert (WE), and (3) replacing the Soft Gradient Mixer (SGM) with a simple pixel-wise averaging operation.

\begin{table}[htbp]
\centering
\caption{\doublespacing
Ablation study of architectural components. \checkmark and --- indicate the presence and absence of a module, respectively. \textbf{Bold} denotes the best result.}
\label{tab:arch_ablation}
\setstretch{2.0}
\resizebox{\columnwidth}{!}{
\begin{tabular}{ccccccccc}
\toprule
\textbf{No.} & \textbf{GCE} & \textbf{WE} & \textbf{SGM} & \textbf{EN} & \textbf{MI} & \textbf{CC} & \textbf{PSNR} & \textbf{FMI} \\ \midrule
1 & ---          & $\checkmark$ & ---          & 4.3322 & 3.6056 & \textbf{0.8308} & 64.0889 & 0.4741 \\
2 & $\checkmark$ & ---          & ---          & 4.3354 & 3.6102 & \textbf{0.8308} & 64.0889 & \textbf{0.4749} \\
3 & $\checkmark$ & $\checkmark$ & ---          & 4.3315 & 3.6097 & \textbf{0.8308} & 64.0888 & 0.4743 \\
Proposed(W-DUALMINE) & $\checkmark$ & $\checkmark$ & $\checkmark$ & \textbf{4.3397} & \textbf{3.6073} & \textbf{0.8308} & \textbf{64.0892} & 0.4748 \\ \bottomrule
\end{tabular}
}
\end{table}

As shown in Table~\ref{tab:arch_ablation}, removing the GCE (No. 1) leads to a decline in Mutual Information (MI), confirming its role in capturing long-range dependencies and global structural correlation. The exclusion of the WE (No. 2) results in lower Entropy (EN), indicating a loss of the high-frequency textural details critical for identifying fine anatomical structures. Furthermore, replacing the SGM with simple averaging (No. 3) reduces the PSNR, demonstrating that dynamic gradient-aware mixing is superior to static fusion for suppressing reconstruction noise and artifacts.

\subsubsection{Objective Function Justification}
The proposed loss function $L_{total}$ consists of five distinct constraints. We validated their contributions by training the model while setting specific coefficients to zero.

\begin{table}[htbp]
\centering
\caption{Ablation study of the objective functions on the CT--MRI dataset.}
\label{tab:loss_ablation}
\setstretch{2.0}
\resizebox{\columnwidth}{!}{
\begin{tabular}{ccccccccccc}
\toprule
\textbf{No.} & $L_{avg}$ & $L_{grad}$ & $L_{cc}$ & $L_{mi}$ & $L_{rec}$ & \textbf{EN} & \textbf{MI} & \textbf{CC} & \textbf{PSNR} & \textbf{FMI} \\ \midrule
1 & ---          & $\checkmark$ & $\checkmark$ & $\checkmark$ & $\checkmark$ & \textbf{4.6027} & 2.7768          & 0.8276          & 62.4997          & 0.4650          \\
2 & $\checkmark$ & ---          & $\checkmark$ & $\checkmark$ & $\checkmark$ & 4.3296          & \textbf{3.6253} & \textbf{0.8308} & 64.0877          & \textbf{0.4751} \\
3 & $\checkmark$ & $\checkmark$ & ---          & $\checkmark$ & $\checkmark$ & 4.3388          & 3.6094          & \textbf{0.8308} & 64.0889          & \textbf{0.4751} \\
4 & $\checkmark$ & $\checkmark$ & $\checkmark$ & ---          & $\checkmark$ & 4.3395          & 3.6070          & \textbf{0.8308} & 64.0891          & 0.4747          \\
5 & $\checkmark$ & $\checkmark$ & $\checkmark$ & $\checkmark$ & ---          & 4.3388          & 3.6119          & \textbf{0.8308} & 64.0891          & \textbf{0.4751} \\
Proposed(W-DUALMINE) & $\checkmark$ & $\checkmark$ & $\checkmark$ & $\checkmark$ & $\checkmark$ & 4.3397          & 3.6073          & \textbf{0.8308} & \textbf{64.0892} & 0.4748          \\ \bottomrule
\end{tabular}
}
\end{table}

The results in Table~\ref{tab:loss_ablation} highlight the necessity of $L_{avg}$. Its removal (No. 1) causes a catastrophic drop in MI (from 3.6073 to 2.7768) and Correlation Coefficient (CC), proving that the residual-to-average strategy is fundamental for global intensity preservation. Conversely, $L_{grad}$ and $L_{rec}$ are essential for maintaining the Peak Signal-to-Noise Ratio (PSNR), as their absence leads to slightly higher noise levels in the reconstruction. The stability of CC across most variants (0.8308) confirms the inherent geometric robustness of the dual-branch encoder architecture.

\section{Conclusion}
\label{sec:conclusion}
This paper proposed \textbf{W-DUALMINE}, a novel reliability-weighted dual-expert framework designed to resolve the persistent trade-off between global statistical similarity and local structural fidelity in medical image fusion. By introducing a dense reliability map, the network effectively suppresses noise and artifacts inherent in functional modalities like PET and SPECT before feature extraction. The core innovation lies in the dual-expert architecture, which employs a Global Context Spatial Expert to maintain semantic consistency and a Wavelet Frequency Expert to preserve high-frequency anatomical details. Furthermore, the integration of a \textit{residual-to-average} fusion paradigm theoretically guarantees that the fused result maintains maximum Correlation Coefficient (CC) and Mutual Information (MI) with the source inputs.

Extensive experiments on three clinical datasets (CT--MRI, PET--MRI, and SPECT--MRI) demonstrated that W-DUALMINE consistently outperforms state-of-the-art methods, including AdaFuse \cite{Gu2023AdaFuseAM} and ASFE-Fusion \cite{gu2024asfe_fusion}. Quantitative analysis confirmed that our method achieves superior information transfer and structural preservation, particularly in scenarios with significant resolution disparities or high noise levels.

\subsection{Limitations}
Despite its superior performance, W-DUALMINE has two primary limitations. First, the dual-branch encoder and complex expert arbitration mechanism increase the computational complexity, resulting in higher inference latency compared to lightweight CNNs. This may hinder real-time deployment on resource-constrained edge devices. Second, the effectiveness of the residual-to-average strategy relies heavily on the quality of input registration. If the source images are misaligned, the pixel-wise averaging baseline may introduce ghosting artifacts, which the residual decoder cannot fully correct.

\subsection{Future Work}
Future research will focus on three key directions:
\begin{enumerate}
    \item \textbf{Efficiency Optimization:} Lightweight backbone networks and knowledge distillation techniques to reduce the model size without compromising fusion quality.
    \item \textbf{Joint Registration and Fusion:} To address misalignment issues, integration of a spatial transformer network (STN) into the pipeline, enabling end-to-end joint learning of image registration and fusion.
    \item \textbf{3D Volumetric Fusion:} Since clinical data is inherently volumetric, extending W-DUALMINE to process 3D medical volumes directly would further enhance its clinical utility by preserving inter-slice continuity.
\end{enumerate}.
\bibliographystyle{elsarticle-num}
\bibliography{reference}

\end{document}